# Lung nodules segmentation from CT with DeepHealth toolkit

Hafiza Ayesha Hoor Chaudhry[1][0000−0001−5932−0103], Riccardo Renzulli[0000−0003−0532−5966], Daniele Perlo[2][0000−0001−6879−8475], Francesca Santinelli[3], Stefano Tibaldi[3][0000−0003−0326−1399], Carmen Cristiano[3], Marco Grosso[3][0000−0002−7083−09295], Attilio Fiandrotti[1][0000−0002−9991−6822], Maurizio Lucenteforte[1][0000−0001−6102−3474], and Davide Cavagnino[1]

[1] University of Turin
[2] Fondazione Ricerca Molinette Onlus
[3] Città della Salute e della Scienza di Torino

**Abstract.** The accurate and consistent border segmentation plays an important role in the tumor volume estimation and its treatment in the field of Medical Image Segmentation. Globally, Lung cancer is one of the leading causes of death and the early detection of lung nodules is essential for the early cancer diagnosis and survival rate of patients. The goal of this study was to demonstrate the feasibility of Deephealth toolkit including PyECVL and PyEDDL libraries to precisely segment lung nodules. Experiments for lung nodules segmentation has been carried out on UniToChest using PyECVL and PyEDDL, for data pre-processing as well as neural network training. The results depict accurate segmentation of lung nodules across a wide diameter range and better accuracy over a traditional detection approach. The datasets and the code used in this paper are publicly available as a baseline reference.

Keywords: Medical image segmentation · Deep Learning · U-Net · Dataset · Chest CT Scan · Lung nodules · DeepHealth

## 1 Introduction and Background

Lung cancer has surpassed breast and prostate cancer, and has become the lead-ing cause of death for men and women in 2021 [22]. Medical Imaging plays a crucial part in the early detection and proper monitoring of cancer patients [11]. Traditionally, a thoracic Computed Tomography (CT) scan of the lungs is per-formed first, which produces high resolution images of the chest structures [18]. Due to the level of detail, great image quality and clear resolution CT scans have become the most popular choice to visualize lung nodules [24]. National Lung Screening Trial (NLST) has also conducted a study that shows the decrease in mortality rate of the lung cancer patients by screening with low-dose CT (LDCT), hence emphasising the role of medical imaging in the recovery process [1][23].

Automated Lung Nodule segmentation done using CT scans could pass es-sential information to the Computer-Aided Diagnosis (CAD) employed for lung cancer diagnosis. A robust lung segmentation method could save the time taken by manual nodule analysis and also remove the inter-observer variability found in many studies [15]. Several CAD systems based on traditional or deep learning image processing techniques have been proposed over the last decade for the de-tection and segmentation of lung nodules [12, 25, 13]. The variations in size and shape of the nodules, the patients' gender and age, the imaging device model and brand, and the resemblance between the nodules and their surroundings make this a challenging problem.

After the detection of lung nodules using medical imaging, the protocol is to have regular follow-up scans from 3 to 12 months, to monitor the growth rate of nodules [14]. To avoid the over diagnosing and to deal with slow growing cancer, the protocols have set the tumor doubling time as an indicator for malignant nodules [10] [20]. To access the tumor response, Response Evaluation Criteria in Solid Tumors (RECIST) is used as a standard, that focuses on the diameter measurement of tumor in uni-direction and linearly [8]. There is a lot of work being done in developing accurate and consistent segmentations of lung tumors for the purpose of response assessment, tumor diagnosis, and staging, which can result in giving linear and volumetric assessments of the tumor size, shape and the tumor change rates.

The largest European lung cancer trail, Nederlands Leuvens Longkanker Screenings Onderzoek (NELSON) trial, focuses on predicting the risk of malignancy in lung nodules. Currently, there are various segmentation algorithms that are automated or semi-automated working on segmentation, detection and classification of lung nodules, but it's difficult to analyse and inter-compare the robustness of them all. A study was conducted using NELSON's data and three different software systems for the malignancy risk assessment task. These systems calculated different tumor doubling time and the results conclude that due to this variation the classification of lung nodules is affected [26].

New lung segmentation algorithms based on learning methods are introduced periodically. Hence there is a dire need of a platform where the user can access different deep learning and computer vision algorithms, analyse them and use them off-the-shelf. This is where our contribution comes in: here we introduce Deephealth toolkit, a complete deep learning and computer vision solution that can be easily used by all developers, includes most commonly available deep learning and computer vision algorithms and provides easy integration of data between them. It also has its own visualisation and image editing tools, therefore providing support from data pre-processing to network fine-tuning.

## 2   DeepHealth toolkit (DHt)

Owing to the success of Artificial Intelligence (AI) and learning-based methods in the health sector, The European Union is funding AI based projects to cater to this challenging field. Deephealth [5, 16] is a health-centered project, part of this effort, that aims to do large-scale experimental research using AI. For this purpose multiple large-scale open access datasets are gathered, UniToChest [17] is one them [2, 17, 9].

The DeepHealth toolkit (DHt) is the framework of Deephealth that is developed to provide one platform for easy deployment of Deep Learning and Com-puter Vision applications [3]. It uses High Performance Computing, Big Data and Cloud Computing for providing off the shelf services for all Image Process-ing related tasks. The Deephealth toolkit consists of European Computer Vision Library (ECVL), European Distributed Deep Learning Library (EDDL), and a front-end interface. In this paper we have used EDDL and ECVL to train deep learning network on UniToChest dataset.

### 2.1  ECVL

The European Computer Vision Library (ECVL) is a general purpose library that is Image centered. It supports basic functionalities like Image read/write, Image manipulation, integration, parallel image augmentation to advance fea-tures like dataset parsing and batch creation. The main objective of ECVL is to provide integration and data exchange between existing Computer Vision (CV), Image processing libraries, and EDDL. Different operating systems including Windows, Mac, and Linux are supported on ECVL. The ECVL contains generic algorithms employed with the Deep Learning library. The Image class in ECVL has a generic tensor model and provides an Hardware Abstraction Layer (HAL) allowing it to run on GPU and FGPU. It also has Memory Management flexi-bility support for different devices. Moreover, there is also a Visualiser for 3D volumes like CT scan slices in ECVL. The user can observe different slices from various views using this visualizer. A basic Image Editor also comes with ECVL, including functionalities like brightness and contrast adjustment, rotation, flip-ping, etc. The programming language of ECVL is C++ but a Python version is also developed using a wrapper class, called the PyECVL, to support Python Ecosystem. The latest versions of ECVL and PyECVL along with documenta-tion are publicly available on GitHub [6] [19].

### 2.2  EDDL

The European Distributed Deep Learning Library (EDDL) is a general purpose library that includes most of the commonly available Deep Learning functional-ities. In addition, EDDL also contains the functionalities that are needed within the Deep Health project, which covers 15 DeepHealth use cases. The main objec-tive of building EDDL is to provide a general-purpose library that covers most of the functionalities needed by Deep Learning in the Health sector, is easy to use and integrate by developers, and becomes the most widely used Deep Learn-ing library for Health. EDDL uses High-Performance Computing (HPC) and a Cloud infrastructure transparently. It provides Neural Network topology compo-nents and hardware-independent tensor operations. Using the Neural Network library both the training and inference of a model can be performed, along with a finer control on deeper levels like gradient manipulation or individual batches. The development language of EDDL is also C++ and a Python wrapper class PyEDDL is provided. The latest versions of EDDL and PyEDDL along with documentation are published publicly on GitHub [7] [19].

## 3   The UniToChest Dataset

The UniToChest dataset has been collected within the EU-H2020 DeepHealth [5, 16] project and consists of more than 300k lung CT scans of pulmonary lungs from 623 different patients. The scans are in DICOM format and each scan comes with a manually annotated segmentation mask in black and white PNG format, both being 512 × 512 in size.

A comparison with similar datasets in Table 1 shows that UniToChest has more nodules with a wider diameter range especially at the top end. The UniToChest contains data collected from a gender-balanced population and span-ning across a wide range of ages. Moreover, it includes images acquired using 10 different devices. The demographic details of patients and insights of the data collection process can be found in the original dataset paper (accepted to ICIAP 2022)[4]. For all the CT scan slices in UniToChest, the radiologist has manually segmented the present lung nodules to provide a segmentation mask. To ensure UE regulation on privacy all the patient identifiers are removed from the CT scan slices and the segmented masks.

| Dataset | Number of Patients | Number of Scans | Total Nodules count | Nodule Diameter range(mm) |
|---|---|---|---|---|
| LIDC − IDRI | 1010 | 244527 | 7371 | 2−69 |
| LUNA16 | 1010 | 888 | 1836 | 3−33 |
| UniT oChest | 623 | 306440 | 10071 | 1−136 |

Table 1: Comparison with similar public dataset shows that our dataset has more clinical lung cancer CT scan slices and annotated lung nodule count with a diverse diameter range.

For the purpose of training a neural network, we split the dataset into training, validation and test set randomly as 80-10-10 of patients. We maintain data consistency across multiple splits by assigning a single split to each patient. The data population with respect to the splits is summarized in Table 2. All the three sets (training, validation and test) have a 60 to 40 ratio between the number of male and female patients. Furthermore, the Table 3 presents an in-depth distri-bution of different nodule diameters within the three splits done for training of the neural network.

| Splits | Number of Patients | Number of Dicoms | Number of Masks |
|---|---|---|---|
| T raining | 498 | 250893 | 18534 |
| V alidation | 62 | 26996 | 1712 |
| T est | 63 | 28551 | 2467 |
| Total | 623 | 306440 | 22713 |

Table 2: Dataset population for the three splits we provide.

|  | < 3mm | < 10mm | < 30mm | > 30mm | Total |
|---|---|---|---|---|---|
| Training | 149 | 6527 | 1861 | 249 | 8786 |
| Validation | 7 | 315 | 116 | 23 | 461 |
| Test | 21 | 575 | 195 | 33 | 824 |
| Total | 177 | 7417 | 2172 | 305 | 10071 |

Table 3: Nodule diameter distribution across three splits.

## 4    Methodology

This section describes the proposed method for pulmonary nodules segmenta-tion, including the preprocessing stage, the architecture of the deep neural con-volutional architecture we rely upon and the relative training procedures. The data preprocessing stage has been accomplished using PyECVL 1.2.0 and for network training PyEDDL 1.3.0 is used.

### 4.1    Data Preprocessing

DICOM files produced by CT machines typically contain pixel intensity values in Hounsfield Units (HU), i.e. they indicate radiometric density per pixel (low val-ues indicating air, higher values indicating bones). Following a standard medical practice, a clipped windowing transformation function is applied to such den-sity values. The window width and center indicate the range of the Hounsfield Units covered inside the converted pixel values, everything outside this range will be equivalent to either zero or one. According to standard practice, we have used a window width of 1600 and a window center of −500 to account for the radiometric density of body structures actually useful for nodule detection.

### 4.2    Network Architecture

Our approach relies on the U-Net implementation [21]. The U-Net consists of Encoder and Decoder part. The encoder consists of 5 convolutional layers with max-pooling for featuremap downsampling. As in other convolutional architec-tures, as the size of the featuremaps shrinks the number of featuremaps increases by a two factor. The decoder includes 5 convolutional layers followed by an up-convolutions, where the size of the featuremaps increases while their number decreases at each layer. A number of encoder and decoder layers are matched with skip connections, where the feature maps generated by the respective en-coder layer is concatenated with the output of decoder layer, enabling the precise learning and localization of image object by allowing different tradeoffs between semantic level and spatial accuracy of the featuremaps. The full architecture of U-Net can be observed in Figure 1.

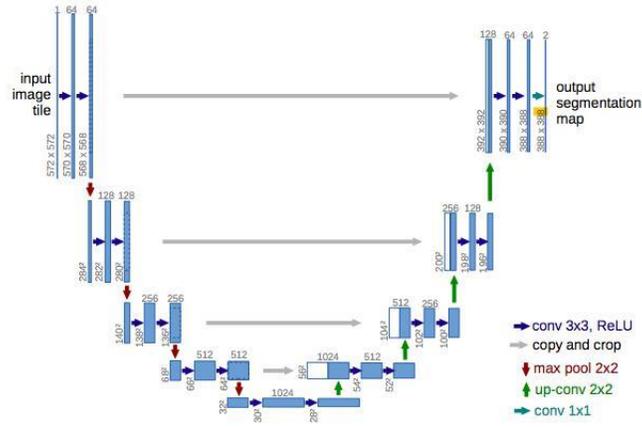

Figure 1: U-Net Architecture[21]

### 4.3   Training Procedure

The training method is fully supervised and consists in randomly initializing the network weights (from scratch) and then training the network for nodule seg-mentation minimizing the loss between the network output and the segmentation mask relative to the input image. As for similar segmentation tasks, we minimize the Dice loss since it has a derivative allowing for error gradient backpropagation and minimizing the Dice loss amounts to maximizing the IoU (Intersection over Union) between predicted and ground truth mask. Next, the network is trained over UniToChest training set for 200 epochs. For this training, only scans with one or more nodules have been considered, since we experimentally verified that other scans do not bring any useful information for segmentation. The CT slices are provided as input to the neural network in batches of 12, as that enabled a reasonable tradeoff between memory footprint and performance. We found beneficial resorting to on-the-fly data augmentation during the training to avoid overfitting to the training data. The augmentation technique we used consists in random flips and rotations (the very same transformations are also applied to the corresponding segmentation mask). The optimizer used in our experiment is Adam with a learning rate of 0.0001. The whole architecture has been implemented in PyEDDL and is available on github.[4]

## 5 Results and Discussion

In this section, we experiment over the UniToChest dataset with the neural network based method described in the previous section for nodule segmentation. All results are relative to UniToChest test set, i.e. images that have not been used at training time. For the experiments Docker image dhealth/pylibs-toolkit:1.2.0-1-cudnn with PyECVL 1.2.0 and PyEDDL 1.3.0 are used.

### 5.1 Experimental Setup

We used Weights & Biases for experiment tracking and visualizations to develop insights for this paper. To automate hyperparameter optimization of the number of workers and the queue ratio size of the ECVL dataloader, we run a sweep with 2 GPUs. In Figure 2 we can see that with the aim of speed up the training process, assigning two workers (one per GPU) and a queue ratio size of eight (four per GPU) lead to better training times. Therefore, when using 4 GPUs we set the number of workers to 4 and the queue ratio size to 16.

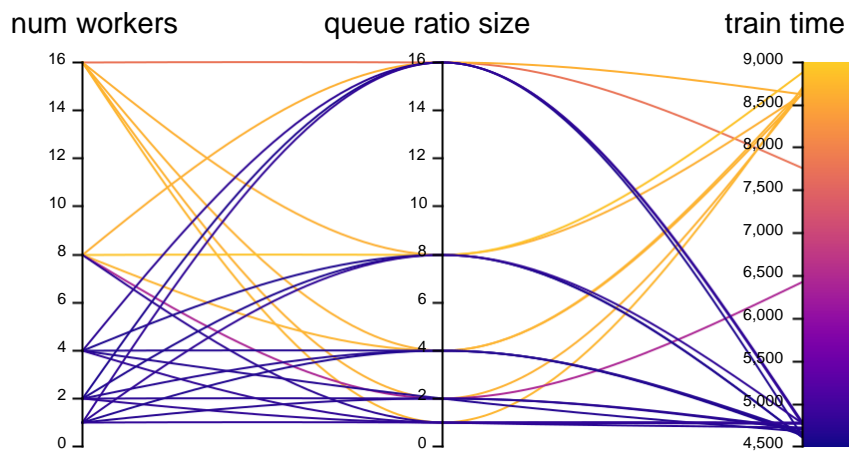

Figure 2: Parallel hyperparameter sweep (grid search) for tuning the number of workers and the queue ratio size of the ECVL dataloader.

---

[4] https://github.com/deephealthproject/UC4_pipeline

### 5.2   Nodules Segmentation

Firstly, for the lung nodule segmentation the neural network was trained using only the images from training split of UniToChest shown in Table 2 that had lung nodules in them i.e. having a respective nodule segmentation mask. The model that performed the best on validation set was picked. This model was then further trained and fine-tuned with 2% of images with black masks, i.e. 2% images that had no nodule segmentation mask, for a few epochs. Lastly, we compute the Dice and IoU scores on all the images of the test set (including images both with and without lung nodules). Figure 3 shows the Dice losses for training, validation and test set when training only on images with nodules. The Dice and IoU scores achieved on the full test set with a model finetuned using 2% of black masks are 0.75 and 0.73 respectively.

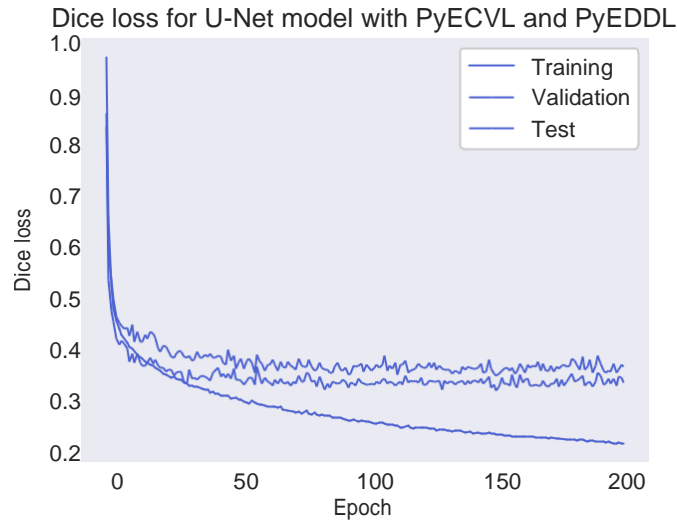

Figure 3: Dice loss when training only on images with nodules.

Finally, Figure 4 shows some samples of the segmentation mask predicted by the network (bottom row) for some sample test images (top row). Red pixels represent false negatives, green pixel false positives and yellow pixels correctly segmented pixels: most of the pixels are correctly segmented, a few errors re-maining only at the borders of the nodule.

### 5.3   Computational speed

DeepHealth libraries provide support for multiple GPUs. We performed experiments using 1, 2 and 4 GPUs with different batch sizes. We can see in Table 4 that running our experiments on bigger batch sizes and using more GPUs we can further speed up training and inference processes.

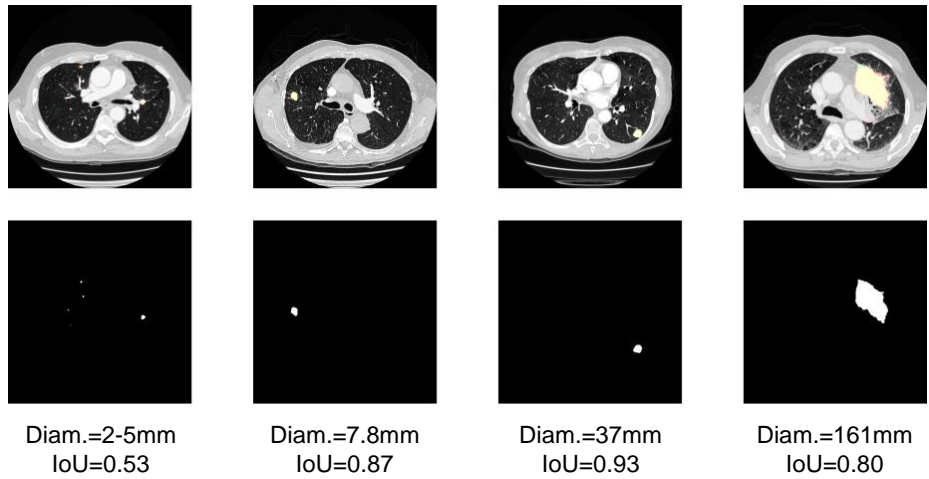

| Diam.=2-5mm | Diam.=7.8mm | Diam.=37mm | Diam.=161mm |
| IoU=0.53 | IoU=0.87 | IoU=0.93 | IoU=0.80 |

Figure 4: Segmentation results on UnitoChest. Overlap (yellow) between pre-dicted (red) and ground truth (green) masks is shown (top). Results over differ-ent nodule diameters and corresponding ground truth are also shown (bottom).

| Number of GPUs | Batch size | Training time (s) | Inference time (s) |
|---|---|---|---|
| 1 | 1 | - | 0.16 |
| 1 | 3 | 6723 | 0.14 |
| 2 | 6 | 4233 | 0.08 |
| 4 | 12 | 3303 | 0.04 |

Table 4: Average training time (seconds, per epoch) and inference time (seconds, per image) with different numbers of GPUs.

## 6 Conclusion and Future Works

This paper proves the feasibility of the Deephealth toolkit and its libraries. In particular, PyECVL and PyEDDL in providing Deep Learning and Computer Vision off-the-shelf services. In this study we proposed a U-Net based architec-ture using PyEDDL that yields promising results at the segmentation of lung nodules from UniToChest. Future research directions of this work include ex-ploiting similar datasets and perform an efficiency comparison with PyTorch.

## Acknowledgement

This work has received funding from the European Union's Horizon 2020 research and innovation programme under grant agreement No 825111, DeepHealth Project.